# Une approche par les modèles pour le suivi de l'activité de construction d'un bâtiment.

## *Bat'iViews* : une interface multi-vues orientée gestion de chantier.

### G. Halin[1], S. Kubicki[2]

[1] *Centre de Recherche en Architecture et Ingénierie*
*2, rue Bastien Lepage - 54001 Nancy, France*
*gilles.halin@crai.archi.fr*
[2] *Centre de Recherche Public Henri Tudor*
*29, avenue JF Kennedy- L1855 Luxembourg-Kirchberg*
*sylvain.kubicki@tudor.lu*

*RÉSUMÉ. La coopération dans les activités de conception et de construction en architecture est un enjeu essentiel dans les mutations organisationnelles et économiques que connaît le secteur. Dans l'état actuel des pratiques professionnelles, de nombreux outils sont utilisés par les acteurs impliqués dans un projet de construction. Les « vues » qu'ils manipulent sont très différentes. Elles ont pour caractéristique de ne représenter que partiellement le contexte de coopération, à travers un point de vue spécifique. « Bat'iViews » propose aux acteurs du chantier une interface de multi-visualisation du contexte et de navigation entre les diverses vues. Cette proposition se fonde sur une approche guidée par les modèles, distinguant la modélisation du contexte de coopération, de la modélisation des concepts représentés dans chacune des « vues-métier » utilisées régulièrement par les professionnels. Une infrastructure d'intégration de ces modèles nous permet d'envisager le développement du prototype et de gérer l'interaction-utilisateur par le biais de transformations de modèles.*

*ABSTRACT. Cooperation between actors in design and construction activities in architecture is an essential stake nowadays. In professional practices the actors involved in construction projects use numerous tools. The project is unique but the "views" that actors manipulate are various and sometimes fundamentally different. Their common characteristic is that they partially represent the cooperation context through a specific point of view. "Bat'iViews" suggests to the actors a multi-view interface of the context and enables to navigate through the different views. This proposition is based on a model-driven approach. We distinguish between "context modelling" and modelling of concepts represented in each "business-view". A model integrative infrastructure allows us to develop the prototype and to manage user interaction through the definition of models' transformations.*

*MOTS-CLÉS : Construction, Chantier, Coopération, Interface de multi-visualisation, Ingénierie Dirigée par les Modèles(IDM), Interface Homme-Machine (IHM), Ingénierie Coopérative*

*KEYWORDS: Building construction, Cooperation, Multi-visualization interface, Model-Driven Engineering (MDE), Human-Computer Interface (HCI), Cooperative Engineering*

**1. Introduction**

Le secteur d'activité de l'Architecture, de l'Ingénierie et de la Construction (AEC[1]) regroupe des acteurs qui sont impliqués dans des actions spécifiques tout au long du cycle de vie du bâtiment (montage de l'opération, conception, construction, réception de l'ouvrage, exploitation et démolition). Dans les opérations de conception et de construction de bâtiments les réseaux d'acteurs impliqués sont *éphémères,* il est donc difficile pour eux de créer des relations durables. Les structures professionnelles sont *hétérogènes* et les « logiques métiers » en situation de coopération sont très différenciées notamment par leurs compétences, leurs modes opératoires, leurs objectifs, et les contraintes liées au métier même ou au type d'entreprise (Evette et Thibault 2000). Enfin, chaque acteur utilise des outils qui sont adaptés aux tâches qu'il doit réaliser, mais aussi à ses compétences ou encore à sa formation.

Si le projet de conception/réalisation est unique, les « vues » que les acteurs manipulent de ce projet à travers leurs outils sont très différentes (Halin 2004). Les intervenants sont *nombreux* et une forte interdépendance temporelle et topologique de leurs interventions rend toujours plus difficile la coordination. Il existe de plus une forte *incertitude* qui est liée à l'activité de production du bâtiment elle-même. Tout bâtiment étant par nature un prototype au sens industriel du terme, la reproductibilité des opérations est donc limitée. La capitalisation de connaissance telle qu'elle s'exprime dans certains domaines de l'ingénierie est presque impossible dans celui du bâtiment. L'activité de réalisation (le chantier) s'effectue évidemment « sur site ». Des problèmes non anticipés apparaissent souvent dus par exemple à la composition géotechnique des sols ou encore aux conditions climatiques de la mise en œuvre (intempéries) (Ward et al. 2004).

Les artéfacts de description du projet sont de nature variée : documents 2D (plans, coupes et façades), maquettes, maquettes 3D (ou virtuelles), pièces écrites (par exemple, les Descriptifs-Quantitatifs-Estimatifs) ou encore images (croquis manuels, simulations d'insertion). En marge de ces documents propres au projet, de nombreux textes réglementaires sont cités en référence (lois, normes, Avis Techniques, Documents Techniques Unifiés pour certains ouvrages particuliers). Enfin, de nombreux documents servent de supports de transmission de l'information de coordination, comme les comptes-rendus de chantier par exemple.

L'échec de l'importation des méthodes dites « d'ingénierie concourante », du domaine de l'industrie au domaine de la construction, n'est pas dû à un retard du monde du bâtiment. Nous faisons l'hypothèse « qu'elles sont fortement inadaptées au contexte singulier de coopération dans le secteur AEC » (Bignon 2002). Dans ce contexte se pose la question de la manière de prendre en compte les spécificités du domaine, que nous avons évoquées. Comme dans tout autre secteur d'activité, les acteurs impliqués dans un travail collectif utilisent des outils afin d'accomplir leurs tâches. Ces outils reposent sur des modèles et des interfaces qui permettent de représenter et d'interagir sur les données manipulées. L'utilisation simultanée de ces différents outils pose un problème, qui n'est pas vraiment résolu à l'heure actuelle : celui de la représentation des données du domaine dans un format interopérable afin de faciliter les transferts d'informations entre ces applications logicielles. Associé à ce problème se pose aussi celui de la variété des interfaces proposées et de leur adaptabilité aux utilisateurs, acteurs du projet, mais aussi aux tâches que ces acteurs ont à effectuer.

---

[1] Ce secteur est souvent appelé AEC, pour « Architecture Engineering and Construction » dans la dénomination anglo-saxonne. Nous ferons régulièrement usage de cet acronyme dans la suite de cet article.

Nous proposons dans cet article, à partir d'une analyse métier du contexte de coopération d'un chantier et de ses dysfonctionnements, de nous intéresser particulièrement aux outils, modèles, et interfaces utilisées au cours de cette activité collective afin de proposer une infrastructure dirigée par les modèles pour la construction d'une interface multi-vues et adaptative aux différents acteurs participant au suivi de l'activité de chantier.

## 2. Le contexte de chantier et ses dysfonctionnements

La figure 1 rend compte, de manière schématique et non exhaustive, de l'organisation d'un chantier particulier[2]. Nous distinguons sur cette figure les intervenants (sur la périphérie), et les dispositifs mis en place en phase chantier pour la coordination : le planning, la réunion de chantier et son compte-rendu.

Le planning prévisionnel est communiqué à tous les intervenants en début d'opération. Il rend compte de la séquentialité des interventions, tout en ne décrivant les tâches qu'à un niveau macroscopique. Sa particularité, dans une activité de chantier, est d'être réactualisé régulièrement pour prendre en compte les changements dans l'activité : aléas météorologiques, retards de livraisons, retards d'exécution etc.

La réunion de chantier est un « outil » de coordination qui se tient de manière hebdomadaire et qui regroupe tous les acteurs du chantier. C'est un lieu d'échanges et de mises au point.

Le compte-rendu a pour objectif de synthétiser les informations échangées au cours de la réunion de chantier et d'assurer leur transmission aux différents intervenants (Grezes et al. 1994). C'est un document de coordination dynamique, actualisé au rythme du chantier. Il est souvent couplé au planning de chantier qui offre une représentation graphique (Gantt, Pert) de l'enchaînement logique des tâches.

---

[2] La recherche présentée trouve une origine et une application dans le suivi expérimental d'une opération de construction d'un établissement public : le collège de Blénod-lès-Pont-À-Mousson (54).

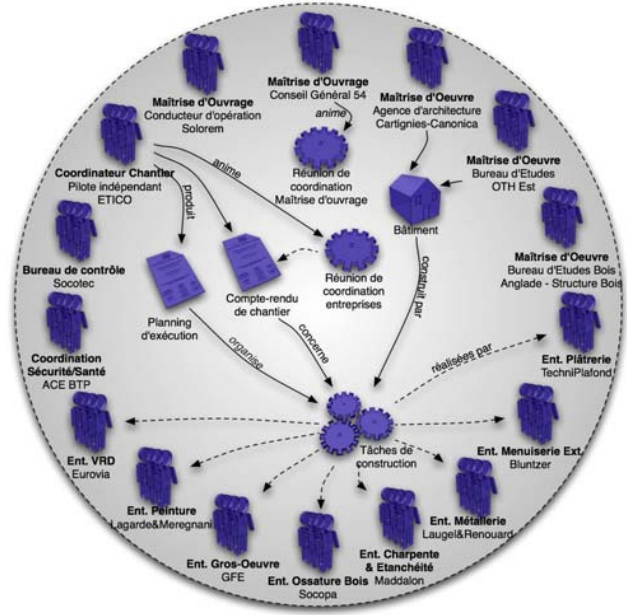

**Figure 1.** Représentation du « contexte coopératif » en phase de chantier

En nous appuyant sur différents travaux (Alluin 1998; Brousseau et Rallet 1995; Hanrot 2003; Tahon 1997), nous distinguerons les dysfonctionnements liés à :

- L'intégration des intervenants : la définition de la mission « suivi de travaux » est bien souvent insuffisante. La répartition des responsabilités et la compréhension du rôle opérationnel de chacun dans le projet est souvent défaillante.

- Les interactions entre intervenants : les processus en cours dans l'activité collective sont peu lisibles et peu partagés. Les acteurs ont peu d'information sur les activités réalisées et surtout ils ont peu de moyens d'en obtenir.

- La diffusion des documents aux personnes concernées et la gestion des versions sont trop régulièrement défaillantes.

- L'activité de construction par elle-même génère aussi des dysfonctionnements. La superposition des tolérances propres à la réalisation de différents ouvrages successifs et à l'intervention de différents corps d'état techniques en est un exemple.

- Les coûts sont relativement difficiles à anticiper et les dépassements de budget sont fréquents. Les problèmes de coordination se traduisent directement ou indirectement par des surcoûts (surfacturation d'un ouvrage non prévu, envois de documents hors délais etc.),

Dans ce contexte, les méthodes de l'ingénierie concourante, classiquement utilisées dans l'industrie, semblent difficiles à appliquer. Plutôt que d'ingénierie concourante nous préférons parler *d'ingénierie coopérative*. L'ingénierie coopérative met en avant un rapport entre les intervenants qui favorise l'ajustement entre les acteurs, un esprit d'équipe et d'initiative pour faire face aux aléas de l'activité et responsabiliser les intervenants en proposant des modalités de coopération (outils, réunions, échanges de documents) adaptées à l'hétérogénéité de ces intervenants. La *flexibilité dans la conduite des processus* est selon nous un enjeu majeur de l'ingénierie coopérative.

Ainsi il semble important que chaque acteur ait une bonne vision du contexte coopératif afin de pouvoir agir et se coordonner avec les autres participants. Se pose alors le problème de la représentation du contexte de l'activité coopérative, et des visualisations que chaque acteur peut avoir de ce contexte relativement au rôle qu'il possède dans l'activité.

Dans l'optique de proposer de nouveaux outils d'assistance à la coordination du chantier, nous allons tout d'abord identifier quels sont les outils actuels ou émergents qui permettent d'assister les activités collectives dans le domaine du bâtiment. Nous proposerons alors une méthode basée sur les modèles pour décrire les outils existants, et définir une infrastructure intégratrice permettant d'offrir une multi-visualisation adaptative aux différents acteurs.

**3. Les outils de coordination du secteur du bâtiment**

Les acteurs du bâtiment utilisent depuis toujours des outils pour réaliser leurs activités particulières. Certains de ces outils, hautement individuels, aident les acteurs à réaliser les tâches qui leur sont attribuées en conception comme en construction.

Durant un chantier, ce sont essentiellement les « outils-documents » de planning et de compte-rendu de chantier décrits plus haut qui sont utilisés. Le plus souvent un simple « outil-logiciel » de traitement de texte ou de tableur est utilisé pour réaliser ces documents.

L'essor des TIC se traduit dans le bâtiment par une adoption de plus en plus généralisée du courrier électronique qui permet aux acteurs de se coordonner par échange de mail et de s'envoyer à moindre coût et plus rapidement des documents volumineux et de meilleure qualité.

L'utilisation des plates-formes de gestion de projet de construction (collecticiels appliqués au domaine du bâtiment) se répand car elles apportent une certaine efficacité dans la conduite des activités collectives. Une étude réalisée par le CSTC[3] (Fassin et Pirlot 2005) montre que seule la gestion documentaire est opérationnelle. Par contre, la gestion de l'activité, impliquant de définir à l'avance et de manière fine les tâches de chacun, est peu pratiquée à l'heure actuelle.

---

[3] Centre Scientifique et Technique de la Construction en Belgique

L'avènement de la CAO et notamment de la modélisation 3D a permis de porter la réalisation d'une maquette dans l'environnement de travail numérique. Mais, la maquette numérique est plus récemment vue comme le modèle du bâtiment permettant l'échange de données et la coopération autour du projet. Ce point de vue a donné naissance au terme anglais « Building Information Model » (BIM).

Le BIM a pour objectif d'unifier la représentation de la connaissance, en distinguant une couche généraliste (le bâtiment lui-même, ses ouvrages décrits), de couches métiers, propres aux actions que les différents acteurs ont à réaliser. En ce sens, le BIM s'écarte donc de la simple maquette, englobant des informations sur les activités coopératives autour de la conception et de la réalisation (matériaux, caractéristiques techniques, performances, produits à mettre en œuvre, aspects normatifs, information de mise en œuvre,…).

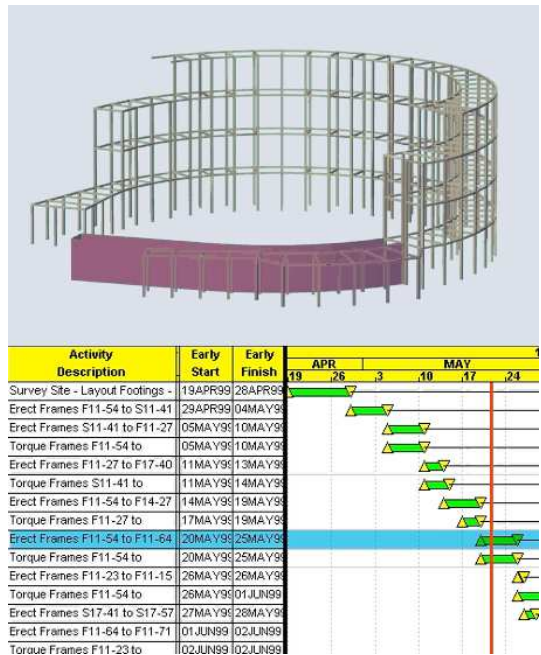

**Figure 2.** Une vue 4D associant planning et modèle 3D

Une autre technologie, récemment issue de la recherche, est de plus en plus expérimentée et utilisée. Il s'agit de la conception 4D des ouvrages (CAO 4D ou 4D CAD, figure 2). La technologie 4D consiste à mettre en relation la maquette tridimensionnelle du projet avec le planning d'exécution des travaux (Chau et al. 2005; Sadeghpour et al. 2004). On obtient ainsi un « planning augmenté » permettant une représentation de l'état d'avancement théorique du chantier dans le temps.

À l'heure actuelle, chaque solution logicielle repose sur un modèle propriétaire de l'information et des services spécifiques qu'elle permet de réaliser répondant à des besoins-métiers hétérogènes (conception, simulations etc.). Des initiatives prônant l'interopérabilité existent cependant depuis de nombreuses années et sont maintenant implémentées dans plusieurs solutions logicielles. L'IAI (International Alliance for Interoperability) développe et fait la promotion du format de données IFC[4]. Deux types d'outils utilisent actuellement ce format :

- Les outils de production (CAO ou autres) propres à différents métiers qui permettent d'importer ou d'exporter des données au format IFC.

- Certaines plates-formes de gestion de projet en ligne permettent le partage de maquette numérique et proposent des échanges avec les différents « logiciels métier » par import-export de tout ou partie de la maquette en ligne. Dans ce type d'outils, la structuration du modèle IFC n'est pas utilisée pour la partie « coopérative ». La plateforme se présente le plus souvent comme un portail de gestion de projet classique. Les connexions entre les objets de la maquette et des tâches ou des documents sont inexistantes dans la plupart des outils.

Les outils que nous avons décrits ci-dessus possèdent une vraie utilité dans la description et la coordination des étapes de la construction. Cependant, il demeure toujours difficile, pour un acteur, d'obtenir une vision globale et synthétique de l'activité du chantier car les outils-documents ou les outils-logiciels présentant cette information sont généralement « sur-renseignés » et « fragmentés ». Si l'anticipation nous semble délicate, nous pensons par contre qu'il est nécessaire dans ces outils de rendre compte de l'état du processus, d'assister la perception de l'activité, afin que chacun puisse adapter son action en fonction de l'état du contexte.

**4. Représentation multi-vues du contexte de coopération : le projet *Bat'iViews***

Le constat à l'origine du projet *Bat'iViews* est le suivant : le contexte de coopération dans une activité de chantier est aujourd'hui représenté par de nombreuses vues attachées à des documents, des outils de coordination ou des outils de communication (figure 3). Pour favoriser la compréhension de ce contexte par les acteurs du chantier, il est nécessaire de trouver une représentation adaptée à chaque acteur-utilisateur mettant en évidence les relations entre les différents éléments de ce contexte.

Notre proposition consiste donc à réutiliser des vues du contexte existantes et manipulées quotidiennement par tous les intervenants du chantier. Leur intégration dans un outil de navigation spécifique nous permettra de rendre explicites les relations entre les divers contenus de ces vues, en proposant à l'utilisateur des interactions sur chacune des vues.

---

[4] Information For Construction : http://www.iai-international.org

Par exemple, des problèmes évoqués dans le compte-rendu de chantier peuvent avoir des répercussions sur l'avancement de certaines tâches. Il existe donc une relation directe entre l'observation dans le compte-rendu et une (ou plusieurs) tâche(s) impactée(s) dans le planning de chantier. De plus, les ouvrages concernés peuvent être représentés dans une maquette numérique.

L'hypothèse que nous formulons est qu'un problème particulier de coordination peut être décrit dans plusieurs vues. À l'heure actuelle, un intervenant du chantier doit convoquer ces vues « manuellement » en croisant des informations dans des documents et des outils hétérogènes. Dans la pratique, cela « coûte » du temps et demande une connaissance de l'ensemble des vues à disposition, donc les acteurs ne le font que très rarement, ce qui entraîne de fréquentes erreurs et incompréhensions.

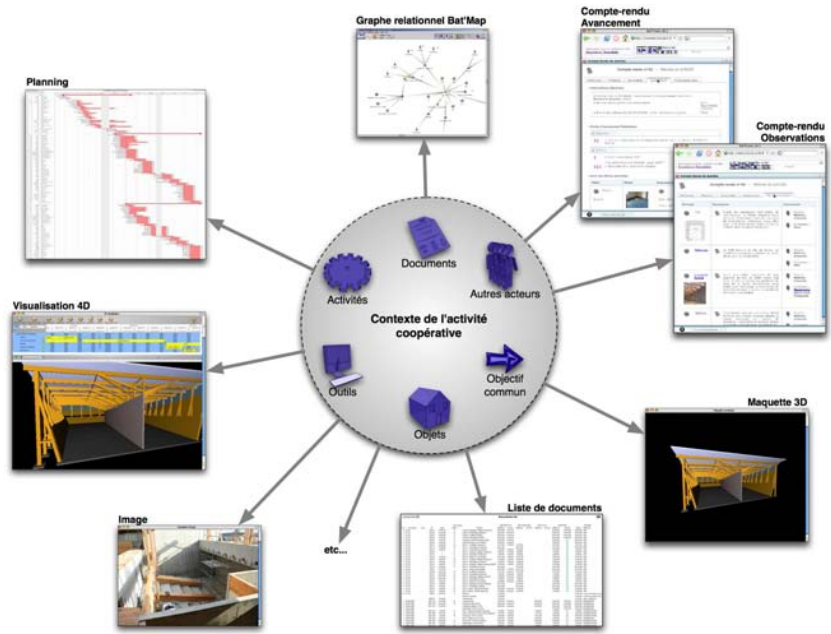

**Figure 3.** Des vues multiples du contexte de coopération

Dans la réalité, de nombreuses vues sont utilisées par les professionnels pour traiter les problèmes de coordination. Nous nous sommes surtout intéressés dans cette proposition aux vues dynamiques : celles qui sont régulièrement actualisées et qui représentent le contexte de coopération sous différents points de vue :

-   La vue du compte-rendu de chantier visualise les problèmes en cours (points particuliers) et l'état d'avancement (informations générales, points d'avancement particuliers, suivi des tâches planifiées).

- La vue planning de chantier décrit l'enchaînement des tâches, d'un point de vue temporel, ainsi que les ressources associées.
- La vue maquette 3D est utile pour convoquer de l'information à propos d'un ouvrage, et représente géométriquement ceux-ci.
- Enfin, une vue représentant les points particuliers de tous les comptes-rendus du chantier sous forme de liste permet de tracer les remarques.

Nous proposons donc de mettre en évidence les relations entre les concepts de chaque vue afin de décrire plus précisément le contexte d'un point de coordination particulier. Pour cela nous nous appuyons sur le principe de la multi-visualisation d'information (North et Shneiderman 1997; Wang-Baldonado et al. 2000) : un arrangement de vues à l'écran permet à l'utilisateur d'avoir une vue globale et de naviguer dans le contexte de coopération en sélectionnant des éléments dans chacune des vues. Lorsqu'un élément est sélectionné dans une des vues, les éléments correspondants dans les autres vues sont alors mis en surbrillance (figure 4).

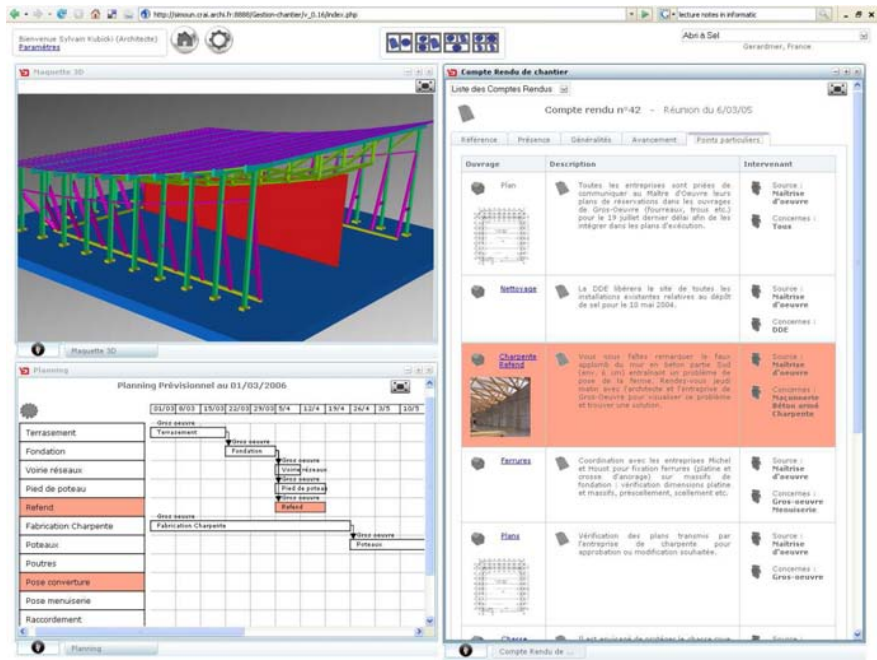

**Figure 4.** Mise en relation d'éléments dans l'interface *Bat'iViews*

Cette proposition vise donc à renforcer deux caractéristiques majeures de la coordination du chantier : la *compréhension de l'information de coordination* par les acteurs (améliorer la qualité de la coordination et sa diffusion) ; et la *conscience de groupe des acteurs* (automatiser la mise en relation de concepts jusqu'ici souvent distincts, et permettre une navigation dans le contexte de coopération à l'aide des vues habituelles).

La juxtaposition de plusieurs vues permet de mettre en relation des éléments de chacune d'entre elles comme le montre la figure 4. Le choix des concepts à relier entre les vues est conditionné par le modèle de chaque vue : le compte-rendu affiche des remarques concernant des ouvrages et des acteurs, le planning affiche des tâches, la maquette 3D affiche des ouvrages. L'interaction est générée par la sélection de l'un de ces éléments dans chacune des vues, et consiste à chercher les concepts correspondants dans les autres vues afin de les mettre en surbrillance. Nous verrons dans la partie suivante que la sémantique liant ces concepts est définie dans un modèle de l'activité coopérative du chantier. *Bat'iViews* (Kubicki 2006) est donc une proposition visant à supporter la « coordination flexible ». Il permet à l'utilisateur de naviguer librement dans le contexte de coopération du projet. Il s'adapte à l'utilisateur en prenant en compte son « contexte acteur » à travers des filtres métiers déterminés à partir de son rôle dans l'activité, et son « contexte utilisateur » en lui permettant de définir des arrangements et d'y introduire des vues particulières. La navigation, quant à elle, a pour objectif de lui permettre de mieux « contextualiser » l'information provenant du contexte de coopération, en mettant en relation les concepts de chaque vue.

**5. Une infrastructure de modèles pour la multi-visualisation du contexte de coopération**

Une branche naissante de la discipline IDM[5] s'intéresse au potentiel des modèles pour la conception des Interfaces Homme-Machine (IHM). De nombreux modèles ont été proposés, pour répondre à différentes perspectives sur l'IHM envisagées par les chercheurs : modèles utilisateur, modèles de tâches, modèles de dialogue, modèles de présentation, architecture logicielle etc. (Bastide et al. 2006).

De nombreuses vues médiatisent le contexte de coopération. Ces vues peuvent êtres « décrites » par un certain nombre de modèles (Bull et Favre 2005; Sottet et al. 2005). Par exemple, l'obtention d'une vue « planning de Gantt » dans un « outil-logiciel » fait appel à différents modèles comme : le modèle du programme (modèle, vue, contrôleur) générant et contrôlant la vue, le modèle du mode de visualisation lié à la technologie employée pour la visualisation (SVG[6] par exemple), le modèle des tâches qui décrit les utilisations possibles de l'interface par l'utilisateur et enfin le modèle des concepts manipulés dans la vue (Tâche, Ressources,..). Ces différents modèles doivent être appréhendés pour décrire la construction et le fonctionnement de cette vue. Dans notre approche, seul le modèle des concepts de la vue (cf. figure 5) qui décrit la sémantique des informations qu'elle manipule est pris en compte.

---

[5] IDM : Ingénierie Dirigée par les Modèles
[6] Scalable Vector Graphics (SVG) est une spécification du W3C. C'est un format de fichier basé sur XML permettant de décrire des ensembles de graphiques vectoriels.

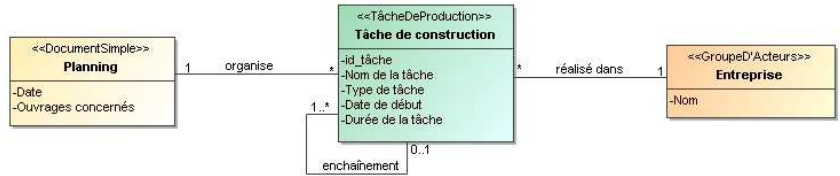

**Figure 5.** Exemple : le modèle de concepts de la vue planning

### 5. 1. Lien entre modèles de concept de vue et modèle du contexte

La « construction » d'une vue fait référence à des concepts issus du contexte de coopération : soit, ces références sont directes (un ouvrage « mur » construit par un « maçon ») soit, elles sont obtenues par une transformation : un « objet3D » dans une vue « maquette numérique » correspond à un « ouvrage » dans le contexte de coopération. Afin de définir ces références entre une vue et le contexte de coopération, nous avons défini une infrastructure permettant l'intégration de modèles pour la construction de vues du contexte de coopération (figure 6). La pyramide schématisant l'infrastructure présente en son axe central les différents niveaux de modélisation du contexte de coopération. Les modèles des vues sur ce contexte gravitent autour de cet axe.

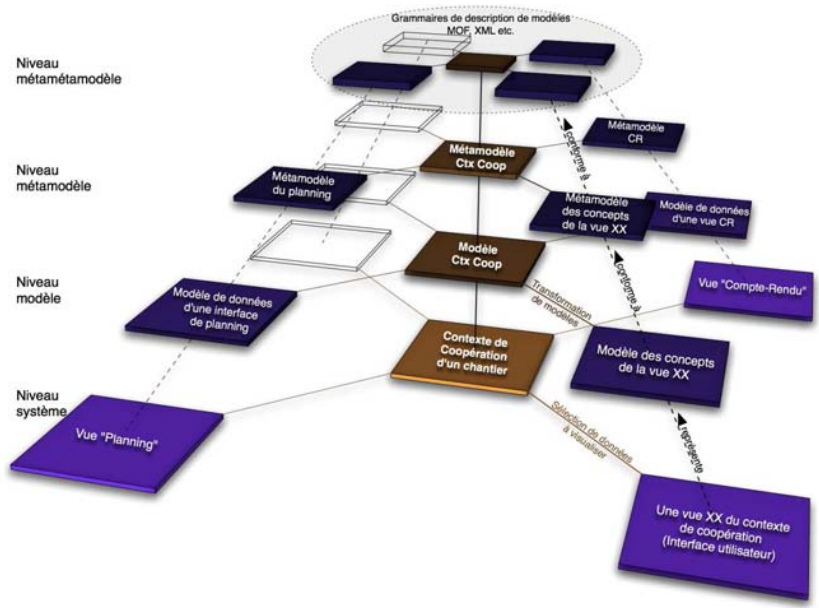

**Figure 6.** Une infrastructure pour l'exploitation des modèles

L'objectif de cette infrastructure est de pouvoir ajouter de nouvelles vues sur le contexte à condition d'en fournir : un modèle des concepts de la vue, et la transformation permettant de construire les concepts de la vue à partir des éléments du contexte. Dans la perspective du développement d'interfaces de multi-visualisation du contexte de coopération, l'unification des modèles que propose cette infrastructure est nécessaire afin d'homogénéiser les relations entre les vues. Ainsi, le modèle du contexte de coopération fournit aux vues la sémantique globale dans laquelle s'insèrent leurs concepts. Cette sémantique permet aux vues d'interagir et ainsi de mettre en correspondance des concepts différents entre deux vues.

*5.2 Modéliser le contexte de coopération*

La description d'une activité collective s'appuie sur un ensemble de concepts de base que l'on retrouve dans de nombreux outils de coopération actuels : « des acteurs sont impliqués dans des activités et manipulent des artefacts ». Ces entités sont à la base des modèles de la plupart des collecticiels disponibles. Cependant chaque activité collective possède une terminologie qui lui est propre (type d'acteurs, type de phases, type de d'artefacts et type de relations).

Les travaux menés au CRAI sur la modélisation de l'activité collective, se sont focalisés sur les activités de conception-construction, qui constituent un contexte particulier de coopération basé sur des relations de nature variable entre acteurs (hiérarchie, ajustement mutuel), activités (interdépendantes ou non) et artefacts (documents et objet à construire).

La proposition repose sur un métamodèle relationnel de coopération qui permet une représentation du contexte coopératif orienté sur la définition des types de relations existant entre les acteurs, les activités et les documents. La figure 7 présente ce métamodèle de coopération. On retrouve les trois entités essentielles d'un contexte de coopération : l'activité, l'acteur, et l'artéfact.

Le rôle de ce métamodèle est de mettre en relation des modèles existant dans toute activité collective. Le concept d'activité fait référence aux modèles de processus, qui sont aujourd'hui très répandus dans de nombreux secteurs d'activités. A titre d'exemple nous citerons le modèle utilisé dans les outils de gestion de flux de tâches (workflow) décrivant les activités et leurs enchaînements dans une activité donnée (WfMC 1999). Le concept d'acteur entretient le lien avec les modèles d'organisation qui permettent de décrire les formes d'organisation d'acteurs impliquées dans une activité collective ainsi que les relations qu'ils entretiennent (Mintzberg 1978). Enfin le concept d'artefact, et plus précisément celui d'objet, fait référence aux modèles de description de produits qui sont utiles à la structuration des données autour des activités de production. Dans le domaine du bâtiment c'est le modèle des IFC qui joue ce rôle. Ce métamodèle n'a pas la prétention de représenter de manière complète la connaissance liée à ces trois domaines. Son intérêt est de permettre la description des relations qui existent entre les concepts de ces différents domaines. La figure 8 présente un exemple d'utilisation du métamodèle pour définir le modèle d'une activité spécifique de construction. Ce modèle a été utilisé pour décrire le contexte de coopération du chantier du collège de Blénod-lès-Pont-À-Mousson présenté dans la figure 1.

### 5.3. Génération du contenu d'une vue et gestion des interactions

La génération du contenu d'une vue particulière, nécessite la définition d'une transformation de modèles pour construire, à partir des concepts du contexte de coopération, les concepts à représenter dans la vue. Cette transformation est définie au niveau des modèles à partir du modèle de coopération (figure 8) et du modèle de concept de la vue concernée (figure 5). Au niveau le plus bas, l'exécution de la transformation opère également une sélection des concepts pertinents à transformer relativement au modèle de concept de la vue, mais aussi en fonction du contexte de l'acteur qui utilise la vue (rôle dans l'activité). La figure 9 illustre ce principe de génération de contenu pour la vue « planning », avec XML comme langage support de description des contenus échangés respectant les XML schémas générés à partir des modèles des concepts des vues.

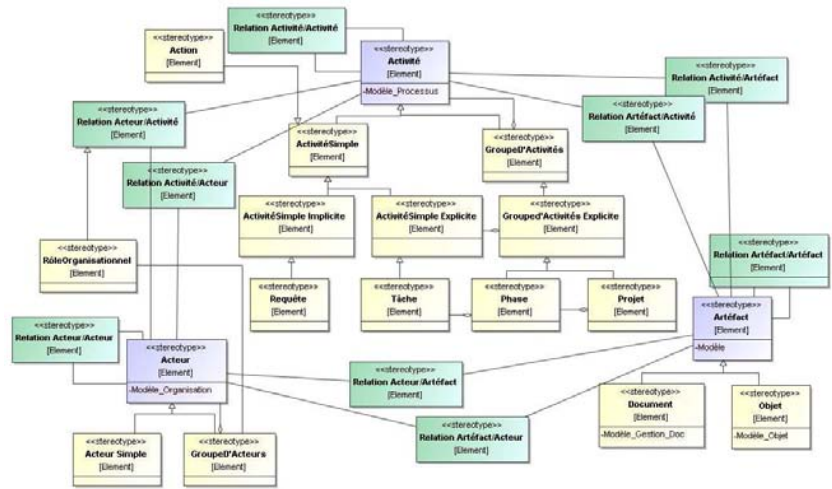

**Figure 7 :** Le métamodèle du contexte de coopération

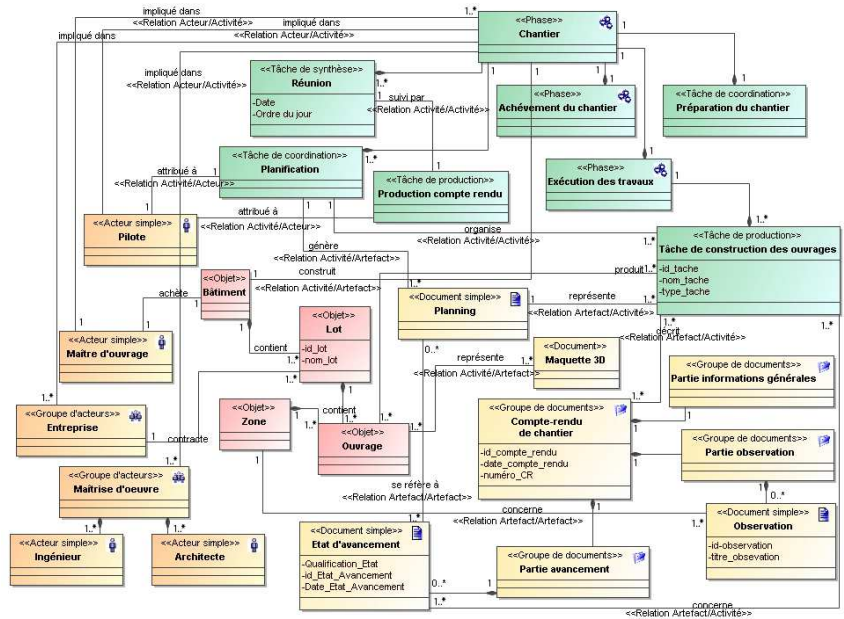

**Figure 8 :** Un modèle de contexte (dédié au chantier) obtenu à partir du métamodèle

La gestion des interactions entre les vues repose sur le paradigme MVC[7] et les transformations de modèles définies entre le modèle de contexte et les modèles des concepts des différentes vues. Chaque vue possède un contrôleur, qui récupère les événements générés par l'utilisateur et accède aux données de la vue (modèle), et il existe un contrôleur général qui coordonne les vues, en récupérant les événements générés par les contrôleurs des vues, et qui exécute les transformations pour accéder aux données du contexte de coopération. Le principe que nous avons appliqué pour définir des interactions entre les vues est le suivant (figure 10) :

1. L'utilisateur sélectionne un élément dans une vue (vue n°1), le contrôleur de la vue identifie l'élément et le concept sélectionné (ici un point particulier sur un ouvrage) dans son modèle,

2. Le contrôleur de la vue génère un événement récupéré par contrôleur du contexte,

3. Le contrôleur de contexte génère un fichier XML qui contient la partie du contexte de coopération en relation avec l'élément sélectionné dans la vue (ici les tâches relatives à l'ouvrage du point particulier). Puis le contrôleur de contexte met en œuvre les transformations de modèle nécessaires pour l'obtention des données (fichiers XML) conformes aux modèles des différentes vues présentes.

---

[7] MVC : Modèle Vue Contrôleur

4. Le contrôleur de contexte génère un événement récupéré par chacun des contrôleurs des vues qui changent le contenu affiché par leur vue (ici le planning).

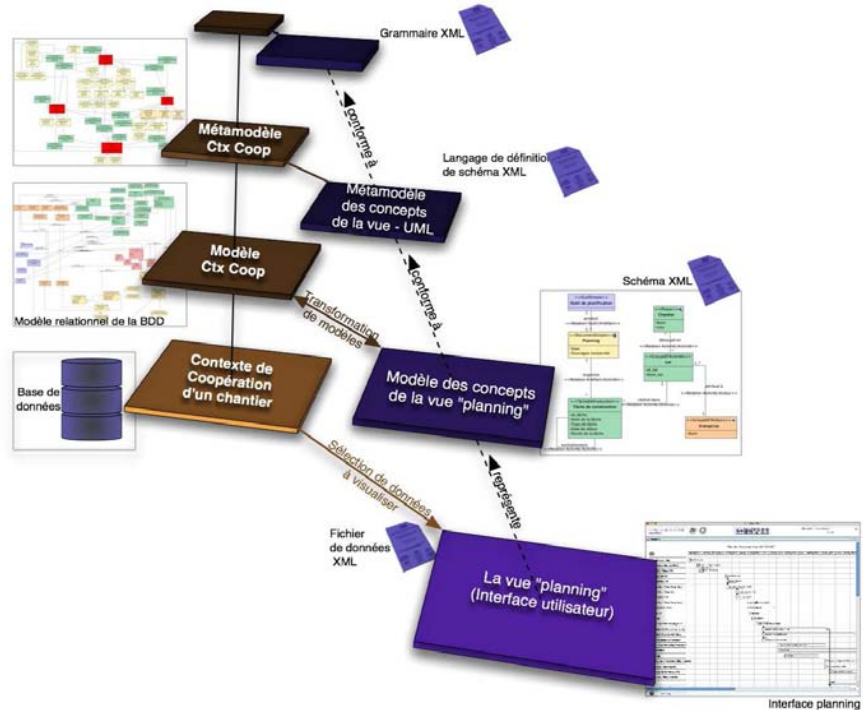

**Figure 9.** Principe de génération de contenu avec la vue planning

Une première version du prototype *Bat'iViews*[8] validant le principe de l'infrastructure proposée a été développée en utilisant une architecture Web trois tiers. Le contexte de coopération est géré par un SGBD (MySQL), les contrôleurs et les transformations ont été programmés en PHP. La partie cliente a été développée en AJAX afin d'optimiser la mise à jour des contenus des vues. Une seconde version est en cours d'étude, elle a pour objectif d'utiliser la technologie MOF QVT (OMF 05) préconisée par l'OMF pour spécifier et mettre en œuvre les transformations de modèles. Les transformations vont être décrites sous forme de règles en utilisant le langage ATL[9] (Bézivin et al. 2003) et sa mise en œuvre dans la plateforme Eclipse. La machine virtuelle d'ATL nous permettra de mettre en œuvre

---

[8] Présentation du prototype : http://www.crai.archi.fr/bativiews, ou démonstration en ligne : http://194.199.221.175:8888/Gestion-chantier/v_0.16/ : login « démo », mdp « demo ».
[9] ATL : Atlas Transformation Language (http://www.sciences.univ-nantes.fr/lina/atl/atldemo/adt)

ses transformations sous la forme de services Web accessibles par le client AJAX de *Bat'iviews*.

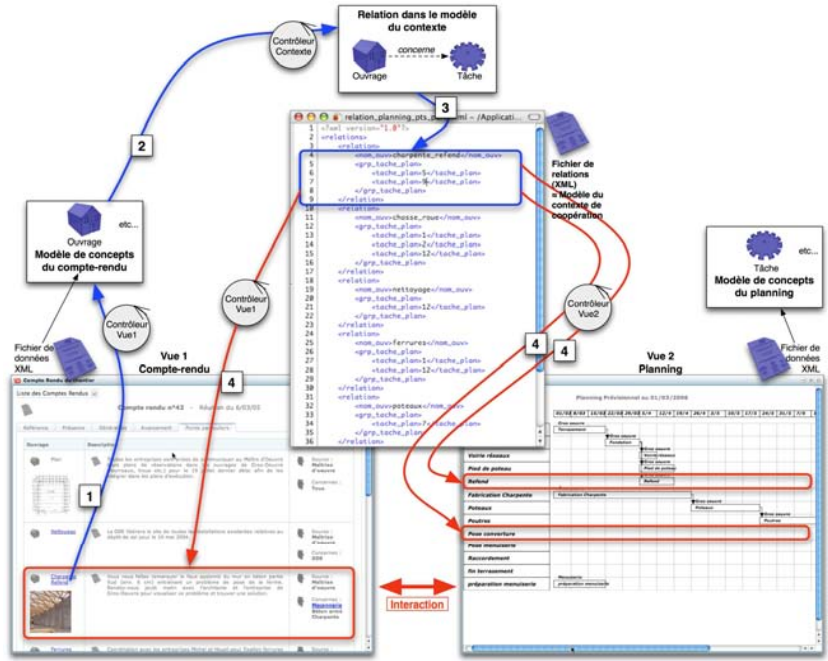

**Figure 10.** Illustration du principe interactionnel mis en oeuvre

## Conclusion

L'activité de chantier, sa coordination et son suivi mettent en œuvre une variété d'outils et de modèles utilisés de manière fragmentée par un ensemble d'acteurs. La mise en correspondance entre les informations générées (le contexte de coopération) par l'activité est rendue difficile par la disparité des modèles guidant cette activité (processus, produit, organisation).

Afin d'apporter une meilleure vision du contexte de coopération à chacun des acteurs, élément essentiel d'une forme *d'ingénierie coopérative*, nous proposons une multi-visualisation du contexte reposant sur une infrastructure de modèles. La « colonne vertébrale » de cette infrastructure est un modèle de coopération que repose sur un métamodèle dont le rôle est de représenter les relations existant entre les différents modèles manipulés dans le suivi de chantier. Cette proposition est rendue fonctionnelle dans le développement du prototype *Bat'iViews*, à travers une intégration des différents modèles de concepts des vues avec le modèle du contexte de coopération et une mise en œuvre des transformations de modèles nécessaires.Il est important de remarquer que la modélisation du contexte proposée est certainement limitée et provisoire. Cette modélisation comble le manque d'un référentiel partagé sous la forme d'une ontologie générale décrivant la sémantique complète du domaine de la construction, rôle que les IFC pourront peut-être jouer prochainement.

**Références**

Alluin P., *Ingénieries de conception et ingénieries de production. L'ingénierie dans les entreprises et industries du bâtiment et ses rapports avec la maîtrise d'oeuvre,* Paris, Plan Urbanisme Construction Architecture, Programme Concevoir. Pratiques de projet et ingénieries, 1998.

Bastide R., Demeure A., Favre J. M., "Atelier IDM & IHM : Ingénierie Dirigée par les Modèles et Interaction Homme-Machine." *IHM 2006, 18e Conférence Francophone sur l'Interaction Homme-Machine.*, Montreal, Canada, 18-21 avril 2006.

Bézivin J., Dupé G., Jouault F., Pitette G., Rougui J. E., "First experiments with the ATL model transformation language: Transforming XSLT into XQuery", *OOPSLA Workshop - 18th annual ACM SIGPLAN Conference on Object-Oriented Programming, Systems, Languages, and Applications*, Anahein, California, USA, October 26-30, 2003.

Bignon J. C., "Modélisation, simulation et assistance à la conception-construction en architecture", *Habilitation à Diriger les Recherches*, Institut National Polytechnique de Lorraine, Nancy, 2002.

Brousseau E., Rallet A., "Efficacité et inefficacité de l'organisation du bâtiment : une interprétation en termes de trajectoire organisationnelle", *Revue d'économie industrielle*, vol. 74, 1995.

Bull I. R., Favre J. M., "Visualization in the Context of Model Driven Engineering", *International Workshop on Model Driven Development of Advanced User Interfaces, MDDAUI 2005 @ MODELS*, Montego Bay, Jamaica, October 2, 2005.

Chau K., Anson M., Zhang J., "4D dynamic construction management and visualization software", *Automation in Construction*, vol. 14, 2005, p. 512-524.

Evette T., Thibault E., *Interprofessionnalité et action collective dans les métiers de la conception urbaine et architecturale,* Paris, Plan Urbanisme Construction Architecture, Cahiers Ramau numéro 2, Rencontres RAMAU des 28 et 29 septembre 2000.

Fassin J., Pirlot D., "Les portails de projet. La gestion collaborative électronique de documents dans les projet de construction", Centre Scientifique et Technique de la Construction, 2005.

Grezes D., Henry E., Micquiaux D., Forgue M., "Le compte-rendu de chantier. Rapport final de recherche", Grenoble, 1994.

Halin G., "Modèles et outils pour l'assistance à la conception. Application à la conception architecturale", *Habilitation à Diriger les Recherches*, Institut National Polytechnique de Lorraine, Nancy, 2004.

Hanrot S., *Enjeux pour l'ingénierie de maîtrise d'oeuvre,* Paris, Ministère de l'Equipement Plan Urbanisme Construction Architecture, Pratiques de projet et ingénieries, 2003.

Kubicki S., "Assister la coordination flexible de l'activité de construction de bâtiment. Une approche par les modèles pour la proposition d'outils de visualisation du contexte de coopération", *Thèse de doctorat*, Université Henri Poincaré, Nancy, 2006.

Mintzberg H., *The structuring of organizations: A synthesis of the research,* Englewood Cliffs, NJ, Prentice-Hall, 1978.

North C., Shneiderman B., "A taxonomy of multiple window coordinations", Human Computer Interaction Lab, University of Maryland. Tech Report HCIL-97-18, 1997.

OMG, "MOF QVT Final Adopted Specification", http://www.omg.org/docs/ptc/05-11-01.pdf, 2005.

Sadeghpour F., Moselhi O., Alkass S., "A CAD-based model for site planning", *Automation in Construction*, vol. 13, 2004, p. 701-715.

Sottet J.-S., Calvary G., Favre J. M., "Ingénierie de l'Interaction Homme-Machine Dirigée par les Modèles", *IDM'05 Premières Journées sur l'Ingénierie Dirigée par les Modèles*, Paris, 30 juin - 1er juillet, 2005.

Tahon C., *Le pilotage simultané d'un projet de construction,* Paris, Plan Construction et Architecture, Collection Recherche n°87, 1997.

Wang-Baldonado M., Woodruff A., Kuchinsky A., "Guidelines for Using Multiple Views in Information Visualization", *AVI,* Palerme, Italy, May 23-26, 2000.

Ward M., Thorpe T., Price A., Wren C., "Implementation and control of wireless data collection on construction sites", *ITcon Electronic Journal of Information Technology in Construction*, vol. 9 Special Issue "Mobile Computing in Construction", 2004, p. 297-311.

WfMC, "The Workflow Management Coalition Specification, Interface 1: Process Definition Interchange Process Model", Workflow Management Coalition, Winchester, United Kingdom, 1999.